\documentclass{DISproc}

\begin{document}
\title{
Diffractive pQCD mechanisms of exclusive production 
of $b \bar b$ dijets and $W^+ W^-$ pairs
in proton-proton collisions}

\author{{\slshape Antoni Szczurek$^1,$$^2$}\\[1ex]
$^1$Institute of Nuclear Physics PAN, ul. Radzikowskiego 152, PL-31-342 Krak\'ow, Poland\\
$^2$Rzesz\'ow University, ul. Rejtana 16, PL-35-959 Rzesz\'ow, Poland }

\contribID{xy}

\doi  

\maketitle

\begin{abstract}
We discuss central exclusive production of $W^+W^-$ pairs in proton-proton
collisions at LHC. Predictions for the total cross section and 
differential distributions in transverse momentum of 
$W^{\pm}$ and $WW$ invariant mass are presented.
We discuss both $\gamma \gamma \to W^+ W^-$ mechanism as well as
a new mechanism of exclusive diffractive production.
The amplitude for the latter process is calculated in the Durham model.
We compare the two (QED and QCD) types of contributions. 
The diffractive contribution is only a small fraction of fb compared to 
the $\gamma \gamma$ contribution which is of the order of 100 fb.
This opens a possibility of searches for anomalous four-boson
$\gamma \gamma W^+ W^-$ coupling due to physics beyond Standard Model.
\end{abstract}

\section{Introduction}

The $\gamma \gamma \to W^+ W^-$ process is interesting reaction to 
test the Standard Model and any other theory beyond the Standard Model.
The linear collider would be a good option to study the couplings of 
gauge bosons in the future.
For instance in Ref.\cite{NNPU} the anomalous coupling in locally 
SU(2) $\times$ U(1) invariant effective Lagrangian was studied. 
Other models also lead to anomalous gauge boson coupling.

It was discussed recently \cite{royon,piotrzkowski,MMN08} 
that the $p p \to p p W^+ W^-$ reaction is a good
case to study experimentally the $\gamma W^+ W^-$ and 
$\gamma \gamma W^+ W^-$ couplings almost at present. 
Only photon-photon contribution for the purely
exclusive production case was considered so far.

Central exlusive production has been recently an active field of
research \cite{Albrow:2010yb}.
The exclusive reaction $pp\to pHp$ has been
intensively studied by the Durham group \cite{Durham}.
This study was motivated by the clean environment and
largely reduced background due to a suppression of $b\bar b$ production
as a consequence of the $J_z$ = 0 rule in the forward limit.
During the conference some results for the $b \bar b$ production
were shown.


In this communication, we discuss exclusive production of $W^+W^-$ pairs 
in high-energy proton-proton collisions.
The original results have been presented recently in \cite{LPS2012}.
The $pp\to pW^+W^-p$ process going through the diffractive QCD
mechanism with the $gg \to W^+W^-$ subprocess naturally constitutes
an irreducible background for the exclusive electromagnetic 
$pp\to p(\gamma\gamma\to W^+W^-)p$ process. We discuss
the contribution of the diffractive mechanism which could potentially 
shadow the photon-photon fusion.

\section{Diffractive mechanism}

\begin{figure}[h!]
\begin{center}
  \includegraphics[width=5cm]{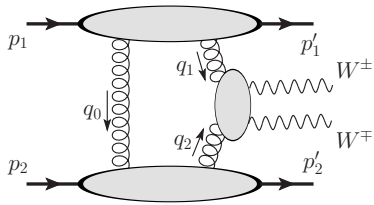}
\end{center}
\caption{
Diagram for the central exclusive diffractive $WW$ pair production 
in $pp$ collisions.}
\label{fig:WWCEP}
\end{figure}

A schematic diagram for central exclusive diffractive production of
$W^{\pm}W^{\mp}$ pairs in proton-proton scattering 
$pp \to pW^{\pm}W^{\mp}p$ is shown in Fig.~\ref{fig:WWCEP}. 

The amplitude of the diffractive process at high energy
is written as:
\begin{eqnarray} \label{ampl}
{\cal M}_{\lambda_+\lambda_-}(s,t_1,t_2) &\simeq&is\frac{\pi^2}{2}
\int d^2 {\bf q}_{0\perp} V_{\lambda_+\lambda_-}(q_1,q_2,k_{+},k_{-})
\frac{f_g(q_0,q_1;t_1)f_g(q_0,q_2;t_2)}
{{\bf q}_{0\perp}^2\,{\bf q}_{1\perp}^2\,{\bf q}_{2\perp}^2}\,,
\end{eqnarray}
where $\lambda_{\pm}=\pm 1,\,0$ are the polarisation states
of the produced $W^{\pm}$ bosons, respectively, $f_g(r_1,r_2;t)$ is
the off-diagonal unintegrated gluon distribution function (UGDF).

The $gg\to W_{\lambda_+}^+ W_{\lambda_-}^-$
hard subprocess amplitude
$V_{\lambda_+\lambda_-}(q_1,q_2,k_{+},k_{-})$
can be formally written as
\begin{eqnarray}\label{GIproj}
V_{\lambda_+\lambda_-}=
n^+_{\mu}n^-_{\nu}V_{\lambda_+\lambda_-}^{\mu\nu}=
\frac{4}{s}
\frac{q^{\nu}_{1\perp}}{x_1}
\frac{q^{\mu}_{2\perp}}{x_2}
V_{\lambda_+\lambda_-,\mu\nu},\quad
q_1^{\nu}V_{\lambda_+\lambda_-,\mu\nu}=
q_2^{\mu}V_{\lambda_+\lambda_-,\mu\nu}=0\,,
\end{eqnarray}
where $n_{\mu}^{\pm} = p_{1,2}^{\mu}/E_{p,cms}$ and the
center-of-mass proton energy $E_{p,cms} = \sqrt{s}/2$.

There are two types of diagrams entering the hard subprocess amplitude:
triangles and boxes \cite{LPS2012}. The corresponding amplitudes
have been calculated using the
Mathematica-based {\tt FormCalc} (FC) package.
The details are explained in \cite{LPS2012}.

The bare amplitude above is subjected to absorption corrections that
depend on the collision energy and typical proton transverse
momenta. The bare cross section is usually multiplied by a rapidity 
gap survival factor which we take the same as for the Higgs boson 
and $b \bar b$ production to be $S_{g} = 0.03$ at the LHC energy. 

The diffractive $WW$ CEP amplitude (\ref{ampl}) described above is
used to calculate the corresponding cross section.
In order to make the calculation feasible we simplify the calculation
limiting to the forward region. The calculation in the full phase space
is obtained by assuming exponential dependence in $t_1$ and $t_2$
and assuming no correlation between outgoing protons. In such an
approximate calculation the phase space integral reduces to four
dimensions \cite{LPS2012}.

\section{Electromagnetic mechanism}

\begin{figure}[h!]
\begin{center}
  \includegraphics[width=7cm]{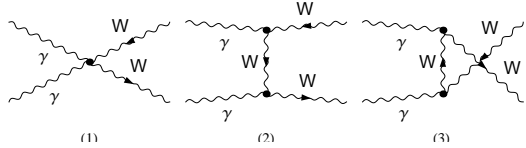}
\end{center}
\caption{The Born diagrams for the $\gamma \gamma \to
W^{\pm}W^{\mp}$ subprocess.}
\label{fig:gamgam_WW}
\end{figure}

In this section we briefly discuss the  $\gamma \gamma \to W^+ W^-$
mechanism. Here we limit to only Standard Model amplitude.
The relevant subprocess diagrams are shown in 
Fig.~\ref{fig:gamgam_WW}. The cross section for the subprocess can
be expressed in terms of Mandelstam variables.

Since we concentrate on the diffractive mechanism the cross section
for the $\gamma \gamma$ mechanism is calculated in approximate way.

%
%

To calculate differential distributions the following parton formula
is used
\begin{equation}
\frac{d\sigma}{d y_+ d y_- d^2 p_{W\perp}} = \frac{1}{16 \pi^2 {\hat s}^2}
\, x_1 f_1^{WW}(x_1) \, x_2 f_2^{WW}(x_2) \,
\overline{ | {\cal M}_{\gamma \gamma \to W^+ W^-}(\hat s, \hat t, \hat u)
  |^2} \, ,
\label{EPA_differential}
\end{equation}
where $x_{1,2}$ are momentum fractions of the fusing gluons.

In our evaluations we use the Weizs\"acker-Williams equivalent photon
fluxes of protons from Ref.~\cite{DZ}.

\section{Results}

\begin{figure}[!h]
\begin{center}
  \includegraphics[width=5cm]{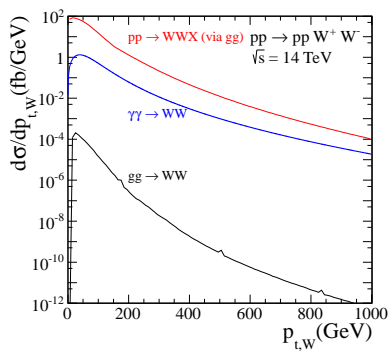}
\end{center}
   \caption{
\small Distribution in transverse momentum of one of the $W$ bosons.
The diffractive contribution is shown by the bottom solid line while
the $\gamma \gamma \to W^+ W^-$ contribution by the middle solid
line. The top solid line corresponds to the inclusive
gluon-initiated $pp \to W^+W^-X$ component.}
 \label{fig:dsig_dpt}
\end{figure}

In Fig.~\ref{fig:dsig_dpt} we show distribution in $W^+$ ($W^-$) 
transverse momentum.
The distribution for exclusive diffractive production is much
steeper than that for the electromagnetic contribution.
The diffractive contribution peaks at $p_{t,W} \sim$ 25 GeV.
This is somewhat smaller than for the $\gamma \gamma \to W^+ W^-$ mechanism
where the maximum is at $p_{t,W} \sim$ 40 GeV.
The exclusive cross section for photon-photon
contribution is at large transverse momenta $\sim$ 1 TeV smaller
only by one order of magnitude than the inclusive $gg \to W^{+}W^{-}$ 
component shown for comparision.

\begin{figure}[!h]
\begin{center}
 \includegraphics[width=5cm]{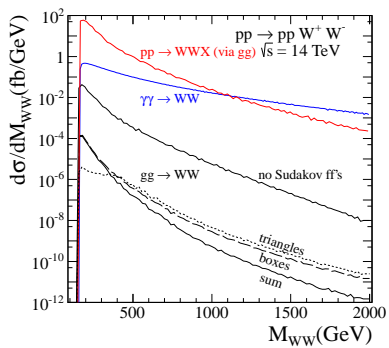}
\end{center}
   \caption{
\small Distribution in $W^+W^-$ invariant mass. We show both the QCD
diffractive contribution and the $\gamma \gamma \to ^+ W^-$
contribution. The result when the Sudakov form factor is
put to one is shown for illustration. The most upper
curve is for the inclusive gluon-initiated $pp \to W^+ W^-X$ component.}
\label{fig:dsig_dMWW}
\end{figure}

Fig.~\ref{fig:dsig_dMWW} shows distribution in the $W^+ W^-$ invariant
mass which is particularly important for the New Physics searches at
the LHC \cite{royon}. The distribution for the diffractive component
drops quickly with the $M_{WW}$ invariant mass. For reference and
illustration, we show also distribution when the Sudakov form
factors in off-diagonal UGDF's is set to one. As can be seen from
the figure, the Sudakov form factor lowers the cross section by a
large factor. The damping is $M_{WW}$-dependent as can be seen by
comparison of the two curves. We show the full result (boxes +
triangles) and the result with boxes only which would be complete if 
the Higgs boson does not exist. At high invariant masses, the
interference of boxes and triangles decreases the cross section even more. 
The distribution for the photon-photon component drops very slowly 
with $M_{WW}$ and at $M_{WW} >$ 1 TeV the corresponding cross section 
is even bigger than the $gg \to W^+ W^-$ component to inclusive
production of $W^+ W^-$.

\section{Summary}

Recently (\cite{LPS2012}) we have calculated the QCD diffractive
contribution to the exclusive $p p \to p W^+ W^- p$ process for the
first time in the literature with the full one-loop $gg\to W^+W^-$
matrix element. The full amplitude is a sum of two mechanisms.
First component is a virtual (highly off-shell)
Higgs boson production and its subsequent transformation into real
$W^+ W^-$ pair. Second component relies on the formation of
intermediate quark boxes.

We have made first evaluation of differential distributions
using amplitudes in the forward limit ``corrected''
off-forward via a simple exponential (slope dependent) extrapolation.
Distributions in $W$-boson transverse momentum and
$W^+ W^-$ pair invariant mass has been presented here for illustration.
The contribution of triangles (with the intermediate s-channel Higgs boson)
turned out to be smaller than the contribution of boxes.
These results have been compared with the results obtained for
purely electromagnetic photon-photon fusion.
We have shown that the diffractive contribution is much smaller than
the electromagnetic one. There are several reasons of the suppression.
Since here we have focused on large invariant masses of the $W^+ W^-$
system, rather large $x$ gluon distributions enter the calculation of 
the diffractive amplitude. The gluon densities at such large invariant
masses, i.e. large $x_1$ and $x_2$, are rather small. Furthermore 
relative to the electromagnetic process the diffractive contribution 
is strongly damped by the Sudakov form factor, soft gap survival
probability and optionally (if Higgs boson exists) by the interference 
of box and triangle diagrams.

In summary, we have given a new argument for the recent idea that
the $p p \to p p W^+ W^-$ reaction is a good place for searches
beyond Standard Model as far as four-boson (anomalous) coupling 
is considered.

\section{Acknowledgements}

I am indebted to Piotr Lebiedowicz and Roman Pasechnik
for collaboration on the issues presented here.
I wish to thank the organizers for perfect organization of DIS2012. 
My stay during the conference in Bonn was supported from a polish grant
DEC-2011/01/B/ST2/04535.




{\raggedright
\begin{footnotesize}



\end{footnotesize}
}


\end{document}